# Cation exchange synthesis of AgBiS$_2$ quantum dots for highly efficient solar cells

Alina Senina[a], Anatol Prudnikau[a], Angelika Wrzesińska-Lashkova[ab], Yana Vaynzof[abc] and Fabian Paulus *[ac]

[a]*Leibniz Institute for Solid State and Materials Research (IFW) Dresden, Helmholtzstraße 20, 01069 Dresden, Germany.*
*E-mail: f.paulus@ifw-dresden.de*
[b]*Chair for Emerging Electronic Technologies, Technische Universität Dresden, Nöthnitzer Str. 61, 01187 Dresden, Germany*
[c]*Center for Advancing Electronics Dresden (cfaed), Technische Universität Dresden, Helmholtzstraße 18, 01069 Dresden, Germany*

Silver bismuth sulfide (AgBiS$_2$) nanocrystals have emerged as a promising eco-friendly, low-cost solar cell absorber material. Their direct synthesis often relies on the hot-injection method, requiring the application of high temperatures and vacuum for prolonged times. Here, we demonstrate an alternative synthetic approach *via* a cation exchange reaction. In the first-step, bis(stearoyl)sulfide is used as an air-stable sulfur precursor for the synthesis of small, monodisperse Ag$_2$S nanocrystals at room-temperature. In a second step, bismuth cations are incorporated into the nanocrystal lattice to form ternary AgBiS$_2$ nanocrystals, without altering their size and shape. When implemented into photovoltaic devices, AgBiS$_2$ nanocrystals obtained by cation exchange reach power conversion efficiencies of up to 7.35%, demonstrating the efficacy of the new synthetic approach for the formation of high-quality, ternary semiconducting nanocrystals.

## Introduction

Metal chalcogenide quantum dots (QDs) have attracted great interest in the last decades due to their exceptional properties such as high colloidal stability, tunable band gap, and high absorption coefficient. These advantages make them highly promising for use in various applications: biological imaging,[1,2] photodetection,[3,4] photovoltaics,[5–7] electrocatalysis,[8,9] *etc*. The area of photovoltaics has seen significant advances over the past 15 years, with quantum dot solar cells reaching high power conversion efficiencies (PCEs), making them of great interest for commercialization.[10,11] However, the most efficient devices are based on lead- and cadmium-based quantum dots, which when produced at mass scale are likely to pose dangerous risks to the environment. These concerns motivate the search for less toxic and heavy metal-free quantum dot materials for photovoltaics. Recent studies have demonstrated that certain binary ($Ag_2S$, $Sb_2S_3$), ternary ($AgInS_2$, $AgBiS_2$, $AgSbS_2$), and quaternary (Zn–Ag–In–S) quantum dots can serve as effective alternatives for solar cell applications.[12–17] Also copper-based material compositions such as ternary (Cu–In–S) and quaternary (Zn–Cu–In–S) quantum dots have been investigated as environmentally-friendly solar cell absorbers, however, these materials require a Grätzel-type solar cell architecture.[18–20] Ternary $AgBiS_2$ QDs have emerged as one of the most promising candidates for application in photovoltaics due to their optimal band gap (1–1.3 eV), high absorption coefficient, ease of synthesis and processing accompanied by high stability in the ambient.

The first attempts to fabricate $AgBiS_2$ based solar cells were made using the sequential ionic layer adsorption reaction (SILAR) technique, demonstrating that $AgBiS_2$ can be used as a potential solar sensitizer.[21] The synthesis of $AgBiS_2$ QDs by hot injection method was initially proposed in 2013 by Chao Chen *et al.*[22] In 2016, Bernechea *et al.* presented the first $AgBiS_2$-based solar cells synthesized by this method with a PCE of 6.3%.[23] Further research was focused on optimizing the synthesis by employing various precursor ratios, changing the injection temperature, and using different capping ligands.[24–26] These attempts in combination with optimized hole and extraction layers resulted in a record result PCE of 9.17%, which further solidified $AgBiS_2$ as a highly promising eco-friendly alternative to toxic chalcogenides.[27] However, despite all the advantages of the hot injection method, it requires both vacuum and high temperatures (typically 100 °C) for several hours, which remains a challenge not only in terms of the thermal budget for QD synthesis, but also the ability to reproducibly synthesize large volumes for commercial applications.

Several studies have explored more inexpensive alternative routes for the $AgBiS_2$ synthesis. For instance, Wu *et al.* proposed a simple synthesis of $AgBiS_2$ nanorods by spin-coating the precursor solutions and subsequent annealing of the films at 300 °C in air.[28] Gu *et al.* demonstrated the synthesis of compact $AgBiS_2$ thin films *via* the formation of Ag-Bi-thiourea-DMSO molecular inks followed by decomposition upon thermal annealing.[29] In addition, Akgul *et al.* presented the first solar cells with $AgBiS_2$ QDs fabricated at room temperature by simple mixing of metal halides with separately dissolved sulfur in oleylamine, resulting in a PCE of 5.55%.[26] These promising attempts highlight the importance of low-cost and low-temperature synthetic routes for $AgBiS_2$ QDs in order to pave the way for their large scale commercial application.

Although direct synthesis allows the rapid preparation of ternary AgBiS$_2$ QDs, this approach is fraught with difficulties due to the presence of precursors of two cations with different reactivities in the reaction system. The different reactivities limit the simultaneous control over size, shape, and composition of the resulting QDs. Cation exchange can be a suitable alternative for a low-temperature synthesis. The cation exchange (CE) reaction is a well-studied method for various material systems in which a guest cation replaces the host cation with subsequent incorporation into the nanocrystal structure.[30] Organic ligands such as phosphines or carboxylates act as soft bases that interact with the softer cation and facilitate its removal from the nanocrystal.[31,32] Modifying the composition of nanocrystals without changing their size and shape makes cation exchange an extremely versatile tool for the synthesis of various materials.

The cation exchange approach has been successfully employed in the synthesis of ternary material compounds.[30,33,34] In the case of CE, easily obtainable binary chalcogenides such as Cu$_2$S, Ag$_2$S, and CdSe are used as a template to further incorporate guest cations (*e.g.*, In$^{3+}$, Hg$^{2+}$) and result in the formation of ternary compounds with precise compositions.[35–37] The outcome of a CE reaction is dictated by subtle thermodynamic and kinetic factors enabling the control over the structure and composition of the desired product.[33] The choice of size, shape, and structure of the parent nanocrystals as well as the choice of solvent, additional ligands, and temperature of the reaction allows not only to tune the morphology and composition of the final products,[38,39] but also to prepare various heterostructures such as Cu–In–Se/ZnS core–shell nanoparticles or CdS/Cu$_2$S binary nanorods.[40,41] Despite the widespread use of CE for synthesizing ternary nanoparticles, the synthesis of AgBiS$_2$ QDs using this approach has not yet been reported.

Here, we demonstrate a facile synthesis of small Ag$_2$S nanocrystals at room temperature and subsequent cation exchange at mild temperatures to form AgBiS$_2$ QDs. The use of bis(stearoyl)sulfide (St$_2$S) as air stable and highly reactive sulfur precursor enables the synthesis of small, almost monodisperse Ag$_2$S nanoparticles at room temperature under ambient conditions. Bismuth neodecanoate (Bi(neo)$_3$) in combination with the soft Lewis base trioctylphosphine (TOP) facilitates the cation exchange of Ag$^+$ to Bi$^{3+}$ at only 50 °C in air. We show that the cation exchange progresses rapidly under these conditions and the size and shape of the nanocrystals does not change throughout the reaction. The AgBiS$_2$ QDs obtained by cation exchange were utilized in planar solar cells that reach a PCE >7%, which is the highest PCE for AgBiS$_2$ QDs synthesized under ambient conditions.

**Results and discussion**

As depicted in Fig. 1, as a first step, crystalline Ag$_2$S QDs were synthesized at room-temperature *via* a quick injection of St$_2$S into a silver nitrate solution in oleylamine (OlAm) and toluene. St$_2$S is a stable, crystalline sulfur precursor that is significantly easier to handle in nanocrystal synthesis compared to the commonly employed hexamethyldisilathiane (TMS$_2$S), which is very sensitive to hydrolysis and air.[42] The use and potential of St$_2$S as sulfur precursor material for the direct synthesis of AgBiS$_2$ and PbS photovoltaic materials has recently been described by us elsewhere.[42] To perform the cation exchange into AgBiS$_2$, purified Ag$_2$S QDs were redispersed in toluene and mixed with trioctylphosphine (TOP) and an excess of Bi(neo)$_3$ was added at 50 °C. Hard soft acid base (HSAB) theory was considered for the prediction of the cations' solubility and their affinity to ligands. According to the HSAB theory, soft Ag$^+$ cations provide a higher affinity to the soft base TOP, allowing for an expedited exchange with harder Bi$^{3+}$ cations.[30,31]

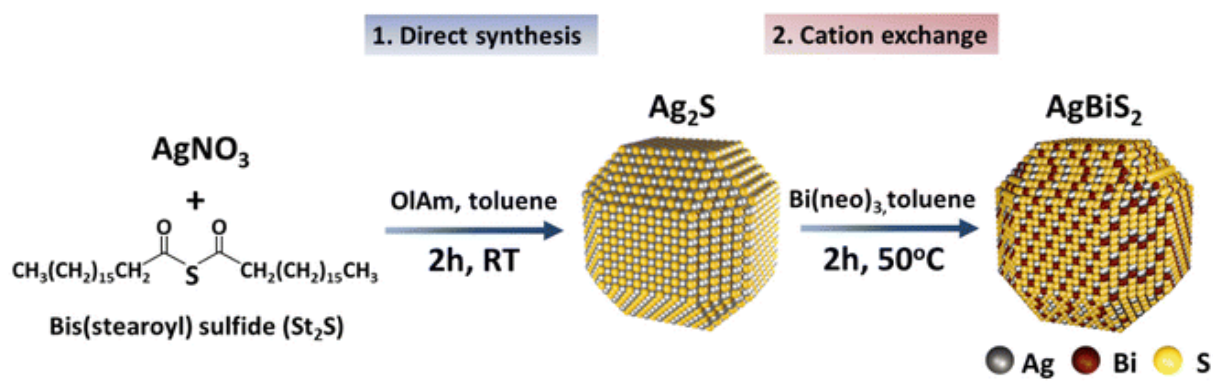

**Fig. 1** Schematic representation of Ag2S QDs synthesis and formation of AgBiS$_2$ via cation exchange at mild temperature.

The kinetics of the cation exchange reaction was monitored using absorbance spectroscopy (UV-VIS), X-ray diffraction (XRD), and transmission electron microscopy (TEM). Fig. 2 shows the TEM images of Ag$_2$S QDs before and throughout the exchange, taken at different times of the cation exchange reaction. The average diameter of St$_2$S-based Ag$_2$S QDs is approximately ≈2.8 nm and exhibits a rather narrow size distribution (see Fig. S1†). Upon the incorporation of Bi$^{3+}$ the average size and shape of the QDs remain the same throughout the reaction. It is also noteworthy that the introduction of TOP into the reaction system does not lead to the full or partial dissolution of the quantum dots, but rather leads to a reduction of larger agglomerates that are present at earlier times of the reaction probably due to a better colloidal stability of the QDs after CE. Consequently, with progressing time of the cation exchange reaction the size distribution becomes narrower (see Fig. 2 and Fig. S1, ESI†).

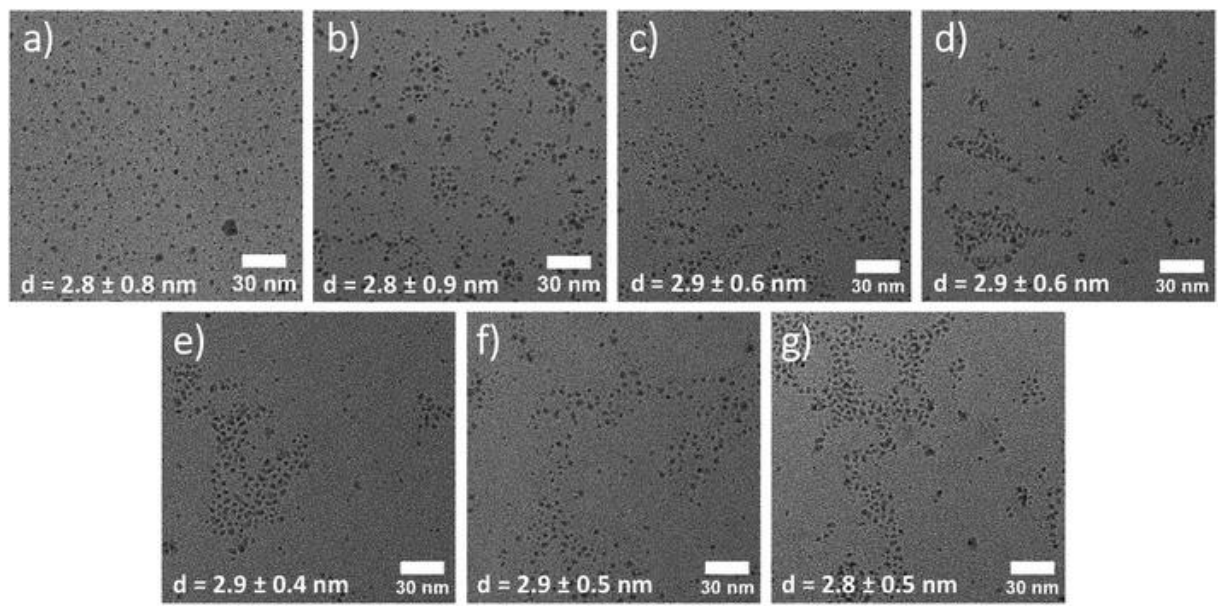

**Fig. 2** TEM images of (a) $Ag_2S$ QDs before and during the cation exchange at (b) 5 min, (c) 15 min, (d) 30 min, (e) 50 min, (f) 80 min, and (g) 120 min of reaction time. Scale bar 30 nm (bottom right) and average QD size and standard deviation (bottom left) for each image.

Fig. 3a shows the evolution of the optical properties of the QDs throughout the cation exchange process. The absorbance of QDs in solution slightly red shifts already after 5 minutes of cation exchange and undergoes only minute changes after 15 minutes. While an absorption shift usually implies a change in the size of the quantum dots for a given material, in this case, it could also be an indication of a change in the band gap. Since bulk, stoichiometric $Ag_2S$ exhibits a bandgap of 0.9 eV, while bulk $AgBiS_2$ exhibits a slightly smaller bandgap of only 0.8 eV, a red shift, as observed, suggests a change in QD composition.[43] However, the shift in bandgap will also be affected by variations of the individual QD stoichiometry and the corresponding exciton Bohr radius of each material, which complicates a full understanding of the observed red-shift. Continuous absorbance measurements at a wavelength of 300 nm throughout the cation exchange process in Fig. 3b support the aforementioned observation that the cation exchange process to $AgBiS_2$ QDs progresses continuously with time and is not based on a full dissolution and new nucleation of nanocrystals. XRD measurements (Fig. 3c) confirm that the transformation from the monoclinic β-$Ag_2S$,[44] to cubic $AgBiS_2$ occurs already within minutes. Subsequent growth in intensity and decrease in peak width for reaction times up to 30 minutes could be an indication of an improvement in the crystallinity of the QDs.

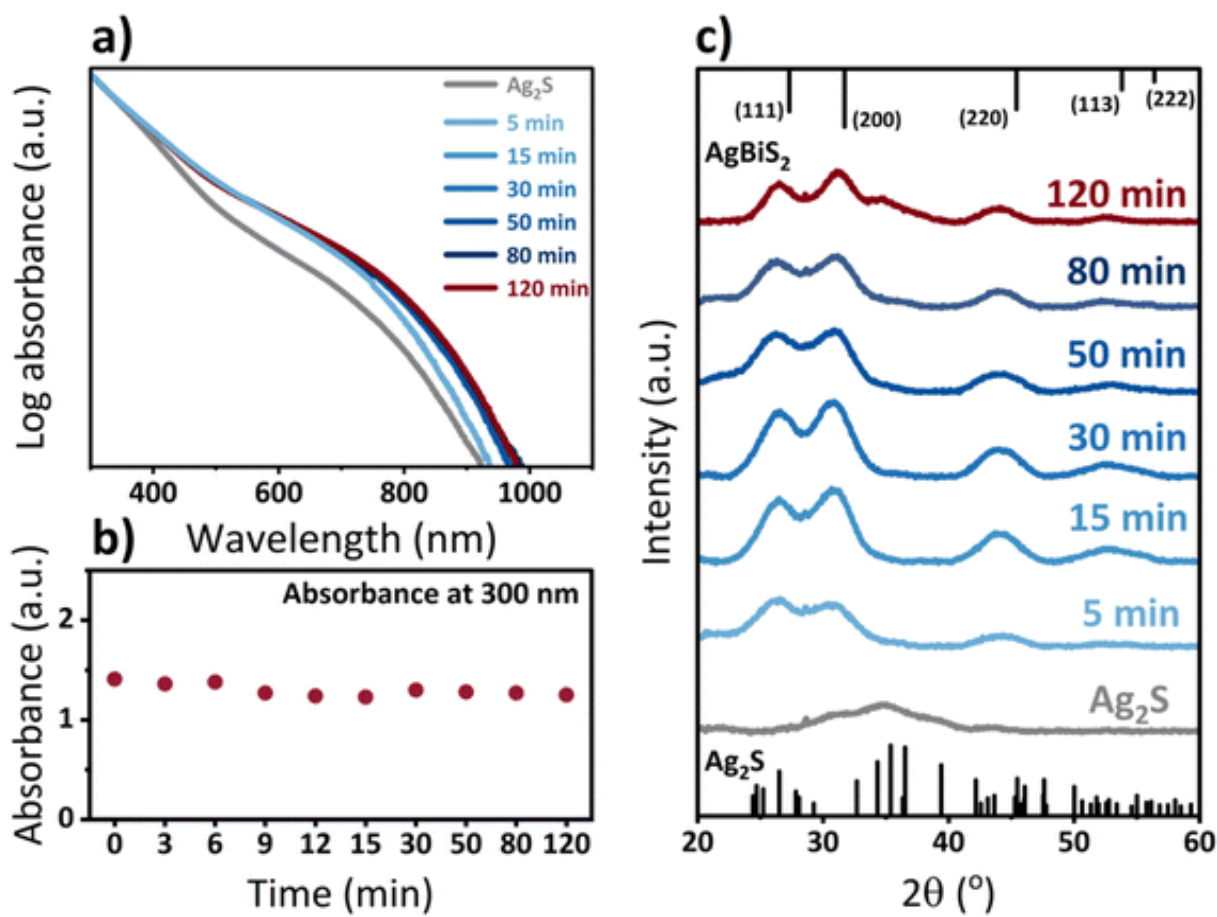

**Fig. 3** (a) Absorbance spectra of $Ag_2S$-based QDs throughout the cation exchange reaction to $AgBiS_2$ at different reaction times, normalized to the absorbance at 300 nm, (b) comparison of absorbance at 300 nm, and (c) XRD patterns of original $Ag_2S$ QDs and $AgBiS_2$ QDs at different times of CE reaction. $Ag_2S$ and $AgBiS_2$ reference spectra in panel c correspond to crystallographic information obtained from ref. 44 and 45 ($Ag_2S$ CCDC 1692248 and $AgBiS_2$ CCDC 1612237), respectively.

To examine the compositional changes of the QDs upon cation exchange, X-ray photoemission spectroscopy (XPS) measurements were performed. The Ag 3d spectra (Fig. 4a) show the presence of a single doublet for both types of QDs, which confirm the absence of metallic silver in the QD films. Comparison of the Ag 3d spectra of $Ag_2S$ and $AgBiS_2$ nanocrystals also reveals the absence of a clear shift in the binding energy upon cation exchange. In the spectrum of S 2p (Fig. 4b), on the other hand, a slight change in the peak position to lower binding energies upon cation exchange is observed, which may indicate a weaker interaction between silver and sulfur after bismuth incorporation. The Bi 4f spectrum consists of two sets of doublets, one associated with $AgBiS_2$ and an additional high binding energy doublet, which is attributed to the formation of bismuth oxide or hydroxylated bismuth species at the QDs surface.[46] The binding energies of the Ag 3d (367.8 and 373.8 eV), S 2p (160.8 and 162.1 eV) and Bi 4f (158.0 and 163.3 eV) doublets that originate from AgBiS$_2$ QDs are in agreement with previously reported data in literature.[23,25,47]

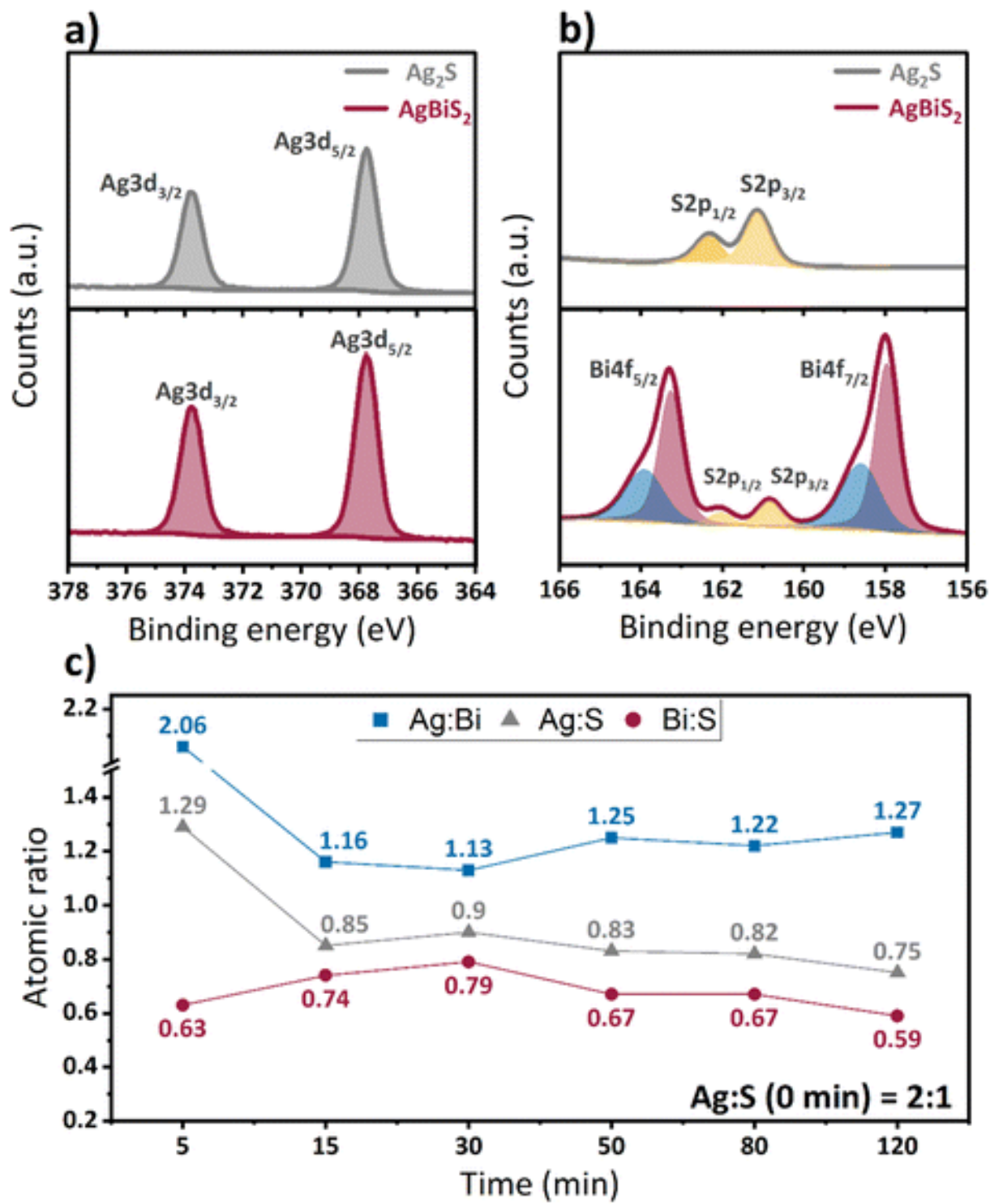

**Fig. 4** XPS spectra of (a) Bi 4f and S 2p, (b) Ag 3d for Ag$_2$S and AgBiS$_2$ QDs after cation exchange of 120 min. (c) Comparison of the atomic ratio between Ag, Bi, and S at different cation exchange times.

The calculated atomic ratios of Ag, Bi, and S in Ag$_2$S and AgBiS$_2$ QDs are shown in Fig. 4c. We observe a sharp decrease in Ag content and an increase in Bi content within the first 30 min of cation exchange, which corroborates a successful exchange of cations. This is followed by a stabilization in the Ag and Bi composition, with a small decrease in bismuth content for very long cation exchange times. The final stoichiometry after 120 minutes is Ag:Bi:S = 0.8:0.6:1 (Table S1, ESI†). While this composition is clearly non-stoichiometric, AgBiS$_2$ QDs synthesized using the hot-injection method using either St$_2$S or TMS$_2$S precursors result in an even larger deviation from the theoretical stoichiometry (Table S2, ESI†). The larger deviation in material composition in the hot-injection method is most likely a result from the formation of non-negligible amounts of silver oleates – and in the case of St$_2$S, silver stearates – in these synthetic procedures, which cannot be easily removed from QDs through the employed redispersion/precipitation procedures.[42] These findings highlight, that synthesis *via* the cation exchange route provides a significantly higher control over the desired QD composition with potentially higher purity.

The synthesis of Ag$_2$S quantum dots using elemental sulfur S and TMS$_2$S as precursors under identical synthetic conditions resulted in the formation of larger Ag$_2$S QDs with sizes of 3.2 ± 1.0 nm and 3.5 ± 1.9 nm, respectively. Both precursors result in more polydisperse nanocrystals (see Fig. S2 and S3†). Attempts to perform a cation exchange for those Ag$_2$S QDs under identical conditions did not result in the formation of pure, high-quality AgBiS$_2$ QDs. For the largest Ag$_2$S QDs no cation exchange could be achieved, indicated by no changes in the diffraction pattern of these nanocrystals (Fig. S2†). While the diffraction pattern changes in the case of Ag$_2$S obtained by elemental sulfur, the absorbance spectrum differs from the one of AgBiS$_2$ obtained from St$_2$S-based Ag$_2$S QDs. Consequently, the integration of QDs based on TMS$_2$S and elemental sulfur as precursors resulted in very low photovoltaic (PV) performance (Fig. S4†). These results could indicate that the Ag$_2$S quantum dot size is crucial for a successful cation exchange with Bi$^{3+}$. To corroborate the size dependence, we synthesized larger Ag$_2$S dots based on St$_2$S. Quantum dots with an initial average size larger than 3 nm do not undergo the cation exchange reaction within 120 min under the same conditions (Fig. S5†). This suggests that only small Ag$_2$S quantum dots (<3 nm) can be converted into AgBiS$_2$ QDs sufficiently fast through cation exchange with Bi$^{3+}$. We note that also other factors may influence the efficacy of cation exchange, such as the choice of ligands, the quantum dot stoichiometry and the presence of defects, however the investigation of these goes beyond the scope of the current study and is subject to future research.

To explore the photovoltaic performance of the QDs formed by cation exchange, they were integrated into solar cells whose architecture is illustrated in Fig. 5a. In short, a thin SnO$_2$ layer was deposited as electron extraction layer (ETL) on top of prepatterned ITO substrates. QDs were deposited in a layer-by-layer process employing a solid phase ligand exchange until a final layer thickness of ≈35 nm was achieved. Next, an ultra-thin poly[bis(4-phenyl)(2,4,6-trimethylphenyl)amine] (PTAA) hole extraction layer (HTL) was deposited.[27] The devices were completed with a thin layer of thermally evaporated $MoO_3$ (that increases the conductivity of the PTAA HTL by p-doping[48,49]) and a Ag electrode. The PCEs and current density – voltage ($J$–$V$) curves of solar cells based on $Ag_2S$ (0 min of CE) and $AgBiS_2$ (30 and 120 min of CE time) are shown in the Fig. 5b and c. It can be seen that while $Ag_2S$ (0 min) demonstrates a very low efficiency, probably due to surface recombination and poor alignment between the ETL and the energy levels of $Ag_2S$. $AgBiS_2$ QDs, on the other hand, exhibit a significant improvement reaching an average performance of 3.31% (best 4.27%) after 30 min of reaction. Interestingly, a further increase in the cation exchange reaction time leads to a stark enhancement in solar cell performance, reaching to an average PCE of 6.76% (best 7.35%). While the open-circuit voltage ($V_{oc}$) is similar for both devices, the fill factor and especially the short-circuit current density are noticeably higher for the QDs after 120 minutes of CE compared to only 30 minutes of CE time (Table S3, ESI†).

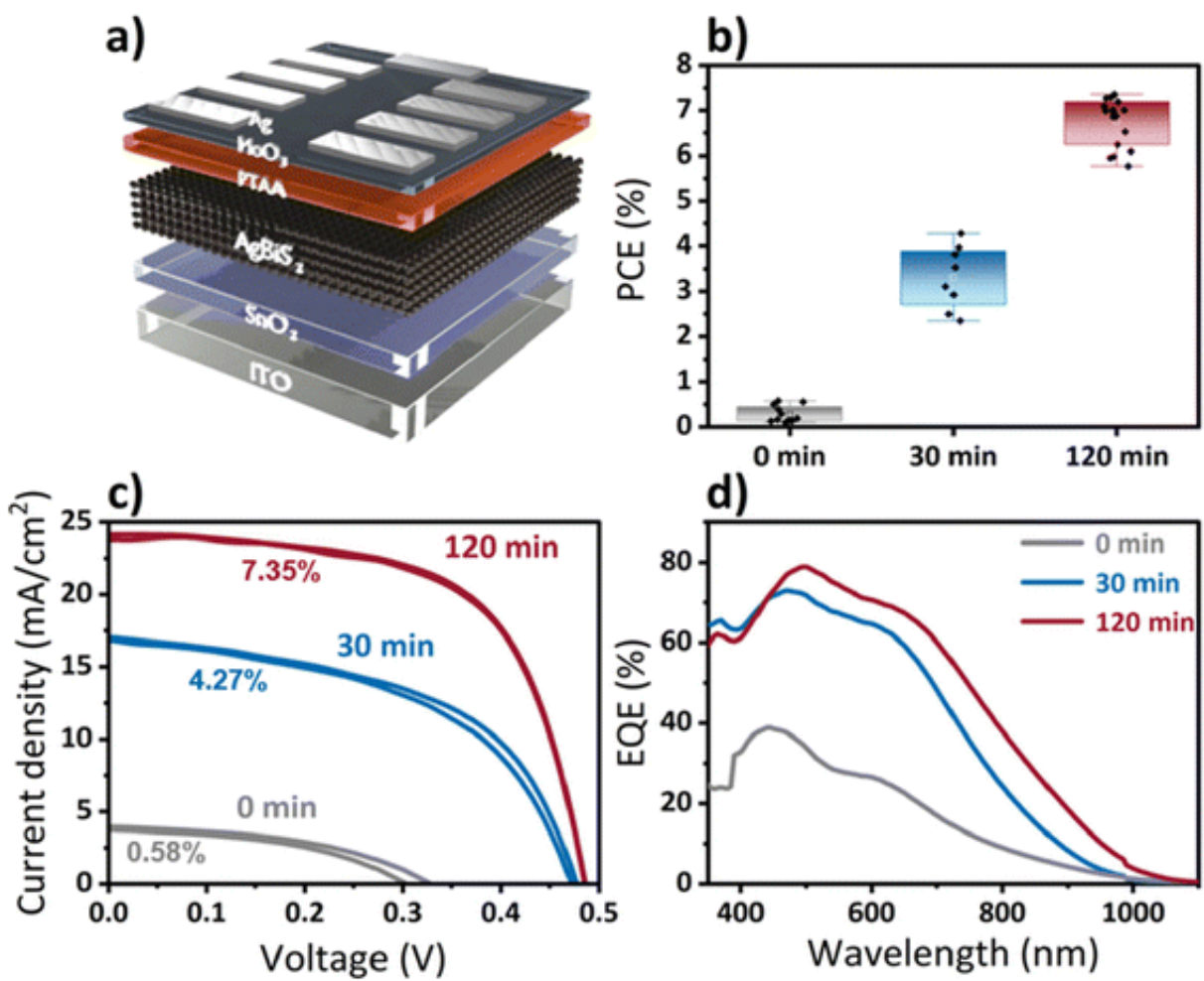

**Fig. 5** (a) Schematic device architecture of AgBiS$_2$ solar cells, (b) Solar cell performance (PCE) of AgBiS$_2$ QDs obtained after different times of cation exchange, (c) *J–V* curves (forward and reverse scanning) of devices with different QDs after 0 min (Ag$_2$S), and AgBiS$_2$ after 30 min, and 120 minutes of cation exchange with individual PCE (d) corresponding EQE curves for these three devices.

Extending the reaction time of the cation exchange beyond 120 min causes the solar cell performance to drop significantly (Fig. S6†). While the exact origin of this decrease in performance remains elusive, the reaction conditions of the cation exchange and the presence of oxidation-sensitive TOP and Ag$^+$-species might affect the AgBiS$_2$ quality, composition or purity for such prolonged reaction times.

We would like to emphasize that the reported solar cell performance represents the performance level of the QD solar cell after one week. Our QD solar cells based on AgBiS$_2$ obtained *via* the cation exchange process exhibit a similar increase in performance within the first few days of storage in ambient as in the case of QDs synthesised *via* the hot-injection synthesis (Fig. S7†).[27] While the exact origin of this improvement remains an open question in the field, our findings suggest that the performance evolution of AgBiS$_2$ solar cell seems to be a universal observation for this device architecture and independent of the synthetic route. Similar observations were also reported for other types of quantum dot based solar cells,[50,51] with the origin of this behavior remaining under debate.

The increase in current density for solar cells of 30 min and 120 min of reaction time is in agreement with the external quantum efficiency (EQE) spectra shown in Fig. 5d, where a maximum EQE of 80% is achieved at 500 nm for the 120 minutes CE devices accompanied by a clear red shift of the EQE onset. Similar changes in the EQE spectra have been attributed in literature to be a consequence of increased cation homogenization within $AgBiS_2$ quantum dots upon thermal annealing.[27,44] We believe that a similar process gradually occurs as the duration of the cation exchange is increased. Cation exchange processes occur through the quantum dot factettes at the surface of the QDs. Within the first minutes of the cation exchange process $Bi^{3+}$ is incorporated into the quantum dot crystal lattice resulting in separated Ag- and Bi-rich domains, while the latter are most likely be in the proximity of the nanocrystal's surface. As the reaction time increases, the overall cation distribution becomes more homogenous across the QD, resulting in a red-shifted EQE, higher short-circuit current and an overall better photovoltaic performance, similar to the observations made by Konstantatos and coworkers,[27] albeit without the need for additional annealing. This simplified model is supported by a detailed examination of the XPS data for bismuth after different times of the cation exchange reaction (Fig. S8†). Higher amounts of bismuth at the nanocrystal surface would result in a higher fraction of hydroxylated or oxidic bismuth species after these QDs have undergone the normal washing procedure and film formation processing, while less of such species are to be expected for bismuth homogenously distributed throughout the entire $AgBiS_2$ QD lattice. Fig. 6 depicts the process of homogenization of the cation disorder schematically and the evolution of the two bismuth species with respect to the reaction time. Our XPS studies show that the ratio between high-binding energy and low binding energy bismuth dramatically decreases by more than 50% with increasing cation exchange duration, while (as discussed above) the overall ratio of the total amount of bismuth to silver remains quasi stable for prolonged reaction times. The hydroxylated or oxidic bismuth species exhibit a doublet at higher binding energies of 158.6 and 163.9 eV as compared to the bismuth of the nanocrystal lattice (158.0 eV and 163.3 eV). These values correspond well to $Bi_2O_3$ and do not stem from the potentially unreacted Bi precursor (Fig. S9†). These data also correlate with previously published XPS spectra of the oxygen-induced degradation of $AgBiS_2$, indicating the formation of $Bi_2O_3$.[46]

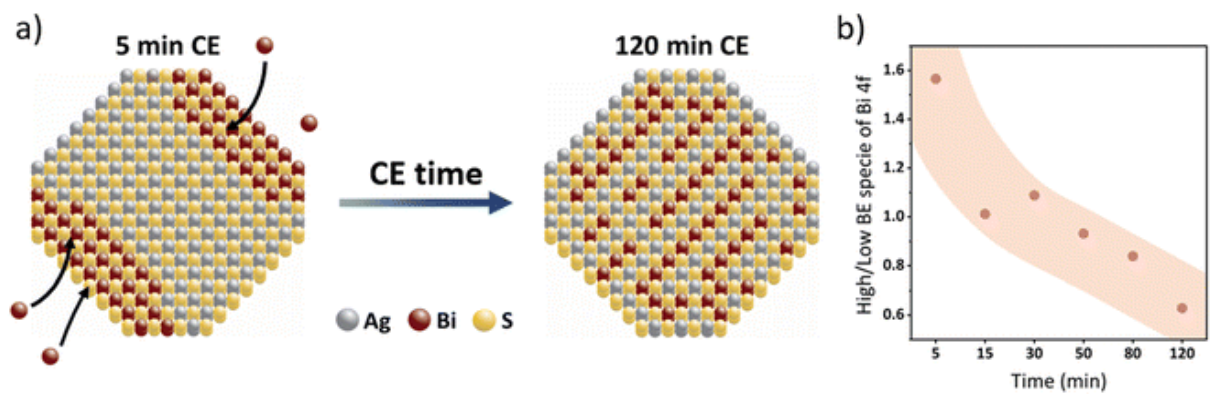

**Fig. 6** (a) Schematic representation of the cation exchange process and resulting homogenization of the cation disorder with increasing reaction time, (b) evolution of the ratio between high-binding/low-binding energy Bi species with CE time.

## Conclusion

In this work, we introduce a cation exchange process as a novel synthetic path for the fabrication of environmentally friendly AgBiS$_2$ QDs at low temperatures and under ambient and vacuum-free conditions. We show that this method enables the preparation of highly monodisperse nanocrystals using bis(stearoyl)sulfide as a non-toxic and air-stable sulfur precursor. The fabricated solar cells exhibit high power conversion efficiency of more than 7% which exceeds previously published PCE value of 5.55% for a ambient, low-temperature synthesis.[26] These results highlight the importance of developing environmentally friendly synthetic routes for the production of efficient QD-based solar cells.

## Experimental section

### Materials

Silver nitrate (AgNO$_3$, 99.5%) was purchased from Grüssing GmbH. Sulfur powder (S, 99.98%), bismuth neodecanoate, oleylamine (OlAm, 98%), mercaptopropionic acid (MPA, 98%), and deionized water were purchased from Sigma-Aldrich. Toluene (99.85%), methanol (99.8%), and acetone (99.8%) were purchased from Thermo Scientific. Trioctylphosphine (TOP, 95%) was purchased from Gelest. Tin oxide (SnO$_2$, 15% in H$_2$O) was purchased from Alfa Aesar. Poly[bis(4-phenyl)(2,4,6-trimethylphenyl)amine] (PTAA, $M_w$ 30 000) was purchased from Ossila. Molybdenum oxide (MoO$_3$, 99.95%) and silver pellets (Ag, 99.99%) were purchased from Kurt J. Lesker. ITO substrates were purchased from Yingkou Shangneng Photoelectric Material Co.

**St₂S synthesis**

Bis(stearoyl) sulfide precursor was synthesized following previously reported procedures.[42,52]

**Ag₂S synthesis**

A silver precursor solution was prepared by mixing AgNO₃ (51 mg, 0.3 mmol) in 2 ml of toluene and 1.5 ml of OlAm. St₂S (85 mg, 0.15 mmol) was mixed with 2 ml of toluene and both precursor solutions were stirred separately for 1 h. After forming a clear solution, sulfur precursor was swiftly injected into the vial with metal precursor resulting in rapid formation of dark brown QDs solution. The reaction system was stirred for 2 h at RT. Then, QDs were precipitated with methanol and centrifuged for 5 min at 6000 rpm. Purified Ag₂S QDs were redispersed in dry toluene. The same synthetic route was performed with TMS₂S and S precursors.

**Cation exchange**

After purification, 24 mg of Ag₂S QDs were dissolved in 2 ml of toluene. Bismuth neodecanoate (0.3 mmol, 230 mg) and 160 μL of TOP were added to the QDs system and mixed at 50 °C for 2 h.

After cation exchange, AgBiS₂ was purified by precipitation and redispersion with acetone and toluene 2 times. Then, QDs were dissolved in hexane and stirred overnight. The solution was centrifuged for 5 min at 10 000 rpm. The precipitate was discarded, the solution was washed with acetone and centrifuged one time. Purified AgBiS₂ QDs were dissolved in dry toluene.

**Device fabrication**

ITO substrates were cleaned in a sonication bath with acetone and isopropanol for 20 minutes and treated with oxygen plasma at 0.4 mbar for 10 min. SnO₂ transport layer was prepared by dilution with deionized water in a volume ratio of 1 : 4. Prepared nanoparticles solution was spin-coated at 2000 rpm for 30 s and then annealed at 270 °C for 15 min. Afterwards, AgBiS₂ or Ag₂S QDs solution with a concentration of 20 mg ml$^{-1}$ in toluene was dynamically spin-coated at 2000 rpm (solution dispensed on a rotating substrate), treated for 45 s with MPA (1% v/v in MeOH), and rinsed twice with methanol and once with toluene. This procedure was repeated three times. Prepared films were transferred into the glovebox and annealed at 115 °C for 10 min. PTAA solution (1.6 mg ml$^{-1}$ in toluene) was prepared and statically spun at 2000 rpm for 30 s. Lastly, 3 nm of MoO₃ and 80 nm Ag were thermally evaporated through a shadow mask with an pixel area of 4.5 mm².

**Measurement and characterization**

**UV-VIS**

A Jasco V-770 spectrometer was used to collect the absorption spectra in the ultraviolet-visible and the near infrared regions. Absorbance measurements were carried out in 1 cm wide quartz cuvettes.

**X-ray diffraction**

Diffraction measurements were carried out on a Bruker D8 Discover diffractometer, with a Cu anode (Kα $\lambda$ = 1.5406 Å) in a coupled Θ–2Θ scan with a 1D-detector. Samples were prepared by drop-casting the nanoparticles solution onto a glass substrate with an area of 1 cm$^2$ without performing any ligand exchange procedures.

**Transition electron microscopy**

A Jeol JEM F200 with an acceleration voltage of 200 kV was used for transmission electron microscopy (TEM) at the Dresden Center for Nanoanalysis (DCN). Carbon grids were used as substrates for drop-cast QDs solutions, heavily diluted in toluene. The average QD size was calculated by measuring the diameters of at least 100 nanoparticles.

**X-ray photoemission spectroscopy**

X-ray photoemission Spectroscopy (XPS) measurements were carried out on an ESCALAB 250Xi by Thermo Scientific in an ultrahigh vacuum chamber (base pressure: $2 \times 10^{-10}$ mbar) with an XR6 monochromated Al Kα X-ray source ($hv$ = 1486.6 eV) and a pass energy of 20 eV. Samples were deposited *via* drop-casting on ITO substrates and annealed at 115 °C for 10 min without performing any ligand exchange procedures. The C 1s state (284.8 eV) was used as a reference to calibrate the binding energy for all presented XPS spectra due to minute differences in surface charging.

**Solar cell characterization**

A Keithley 2450 source measure unit and Abet A+++ solar simulator with an AM 1.5 filter were used for current–voltage measurements. External quantum efficiency (EQE) measurements were performed on a self-built setup with a halogen lamp, monochromator, beam splitter and a calibrated photodiode and filters. The intensity of the solar simulator illumination was corrected in accordance with a calculated mismatch factor from the EQE measurements and calibrated with a silicon reference cell (NIST, traceable) to 1 sun (100 mW cm$^{-2}$). Solar cells were measured each day, starting at day 0, directly after fabrication. During the measurements, the samples were stored in ambient conditions (20 °C < $T$ < 23 °C; 30% < R.H. < 65%) in a cabinet. The performance increases within the first few days, reaches a maximum after 4–5 days and remains roughly stable before slow degradation sets in after around 21 days (Fig. S7†). This behavior is common for quantum dot based solar cells.[50]

**Author contributions**

A. S. conducted the cation exchange synthesis, characterization of QDs and fabrication of solar cells. A. P. provided the St$_2$S precursors and performed TEM analysis. A. W.-L. performed XPS measurements and analysis together with Y. V., F. P. conceptualized and supervised the project. All authors contributed to the writing, editing and commenting of the manuscript.

**Acknowledgements**

A. S., A. P. und F. P thank the German Federal Ministry for Education and Research (BMBF) for funding through the project "GreenDots" (FK 03XP0422A). A. W.-L. and Y. V. received funding from the European Research Council (ERC) under the European Union's Horizon 2020 research and innovation programme (ERC Grant agreement no 714067, ENERGYMAPS). F. P., A. P., and A. S. thank the DCN for free access to the TEM facilities.